\documentstyle[letter,mbf]{ptptex}

\preprintnumber[3cm]{
KUNS-1325\\ HE(TH)~97/04\\ hep-th/9702083}

\title{Cosmological Spinning Multi-`Black-Hole' Solution in String Theory}

\author{Tetsuya {\sc Shiromizu}}

\inst{DAMTP, University of Cambridge \\ 
Silver Street, Cambridge CB3 9EW, United Kingdom\\
\vskip 5mm
Department of Physics, The University of Tokyo, Tokyo 113-0033, 
Japan \\
and \\
Research Centre for the Early Universe(RESCEU), \\ 
The University of Tokyo, Tokyo 113-0033, Japan}

\recdate{
}

\abst{We give a cosmological spinning multi-`black-hole' solution in the 
Einstein-Maxwell-Dilaton-Axion theory with a positive cosmological 
constant. This solution is the cosmological dilatonic 
Israel-Wilson-Perjes solution and describes the collision of several 
spinning `black-holes'.}

\begin{document}

\maketitle


A cosmological multi-black hole solution has been discovered by 
Kastor and Traschen\cite{rf:KT}(KT solution) in the Einstein-Maxwell 
theory with a positive cosmological constant $\Lambda$(EM-$\Lambda$). 
The solution describes collisions of several 
black holes(BHs) in asymptotically deSitter space-times. 
It is worth noting  that the solution is the first exact 
solution such that describes BH-BH collision. It also provides the 
test of the cosmic censorship\cite{rf:Brill,rf:Nakao} and cosmic
no-hair conjectures\cite{rf:nohair} in the inflationary universe
\cite{rf:inf}. 

One may be greatly interested in the spinning version of the KT solution 
because the rotating cases are generic. Apart from the cosmological
cases, the spinning multi-soliton solution was found by 
Israel~\&~Wilson\cite{rf:IW} and Perjes\cite{rf:Perjes}(IWP
solution) in asymptotically flat space-times. 
Unfortunately, the solution has naked singularities due to 
the force balance condition $Q=M$. On the other hand, we can expect
that there are cases without singularities in asymptotically deSitter 
space-times even if $Q=M$. This expectation relies on the causal structure 
of the single Kerr-Newman-deSitter space-time. 
So we can obtain the physical advantage 
of the cosmological version because of no singularity. 

The existence of the spinning multi-soliton solution is deeply related to the 
force balance up to the gravitational spin-spin interaction. The force
cancellation has been confirmed for the IWP solution\cite{rf:Force}. 
The present author and Gen also showed that the force cancellation holds 
between a probe particle and the Kerr-Newman-deSitter 
space-time\cite{rf:Gen}. This observation strongly encourages us to find a 
new cosmological spinning multi-black-hole solution, that is, the 
spinning version of the KT solution.    
 
In this Letter, however, we will give an exact solution in the four 
dimensional Einstein-Maxwell-Dilaton-Axion(EMDA) 
theory with a positive $\Lambda$(EMDA-$\Lambda$) which describes  
a low energy string theory except for a positive $\Lambda$. 
The construction of the solution 
is motivated by the chiral null model\cite{rf:Chiral} in five dimensions. 
In the four dimensional EMDA theory without the cosmological 
constant, there is the dilatonic IWP
solution\cite{rf:SUSY} where the force cancellation
holds\cite{rf:Tess1}. Via a certain dimensional reduction the solution turns
out to be embedded into the chiral null 
model in five dimensions\cite{rf:Chiral}. 
Therefore we realise that it is better that we 
think of an exact solution in five dimensions and we are bearing 
the chiral null model spirit in mind. By virtue of 
the spirit, it is more easy to 
handle the EMDA-$\Lambda$ theory than the EM-$\Lambda$ theory. 

In the Einstein frame the action of the four dimensional 
EMDA-$\Lambda$ theory is 
%
\begin{eqnarray}
S_4 & = & \int d^4x {\sqrt {-g}}\Bigl[R-2(\nabla \phi)^2
-e^{-2\phi}F_{\mu\nu}F^{\mu\nu}
-\frac{1}{12}e^{-4\phi}H_{\mu\nu\sigma}H^{\mu\nu\sigma} \nonumber \\
& & ~~~~~~~~~~~~~~~~~~~~~~~~~~~~~~~~~~~-e^{2\phi}\Lambda
\Bigr]\label{eq:action1},
\end{eqnarray}
%
where $\phi$, $F_{\mu\nu}$ and $H_{\mu\nu\sigma}$ are the dilaton,
field strength of the electromagnetic and three form tensor fields, 
respectively.  In this paper, $\Lambda$ is assumed to be a positive 
constant. 

Firstly, we give the expression of the exact solution without the 
explanation of the derivation. The metric 
$g_{\mu\nu}$, vector potential $A_\mu$ and the three form $H_{\mu\nu\sigma}$ 
are 
%
\begin{eqnarray}
g_{\mu\nu}dx^\mu dx^\nu = a^2\Bigl[-F(d\tau+\omega_idx^i)^2+ F^{-1}\delta_{ij}
dx^i dx^j \Bigr] \label{eq:ans1}
\end{eqnarray}
%
%
\begin{eqnarray}
A=\frac{F}{{\sqrt {2}}}(d\tau +\omega_i dx^i)
\end{eqnarray}
%
and
%
\begin{eqnarray}
H_{\mu\nu\sigma}=24A_{[\mu}\partial_{\nu} A_{\sigma]},
\end{eqnarray}
%
where 
%
\begin{eqnarray}
F^{-1}=1+{\rm Re}\Bigl[
\sum_i \frac{M_i}{a^2|{\mbf x}-{\mbf x}_i|} \Bigr]\label{eq:functionf}
\end{eqnarray}
%
%
\begin{eqnarray}
\omega={\rm Re}\Bigl[
\sum_i \frac{N_i}{a^2|{\mbf x}-{\mbf x}_i|}\Bigr] \label{eq:omega}
\end{eqnarray}
%
and the relation between $\omega_i$ and the above harmonic 
function $\omega$ is given by 
%
\begin{eqnarray}
\epsilon_{ijk}\partial_j\omega_{k}=\partial_i \omega.
\end{eqnarray}
%
$M_i$ and $N_i$ are real parameters. ${\mbf x}_i$ are 
complex constant vectors. The dilaton field is 
%
\begin{eqnarray}
e^{2 \phi}=\frac{F}{a^2}.
\end{eqnarray}
%
$a$ is the function of the time coordinate $\tau$ so that 
%
\begin{eqnarray}
a(\tau)=e^{H_0\tau},\label{eq:ans2}
\end{eqnarray}
%
where $H_0 :=\pm{\sqrt \Lambda}/2$. This solution apparently induces 
the non-cosmological dilatonic IWP solution\cite{rf:SUSY} 
in the limit of $H_0=0$. It also does the non-rotating 
dilatonic KT solution\cite{rf:HH,rf:MS} in the limit of $N_i=0$ and 
the real ${\mbf x}_i$. 

We can directly check that Eqs. (\ref{eq:ans1})$\sim$(\ref{eq:ans2}) 
satisfy the field equations derived from the action of Eq. (\ref{eq:action1}). 
However, as we said, our construction is inspired by the chiral null model 
in five dimensions. So we will guide our trace here. 
The action(Eq. (\ref{eq:action1})) can be obtained from the 
five dimensional theory as follows. 
The five dimensional low energy string action is given by 
%
\begin{eqnarray}
S_5=\int d^5x {\sqrt {-G}}\Bigl[ R_G+4\nabla_M \phi \nabla^M \phi 
-\frac{1}{12}H^2-\Lambda \Bigl] e^{-2\phi},\label{eq:action2}
\end{eqnarray}
%
where $M=y,0,1,2,3$, $H^2=H_{IJK}H^{IJK}$ and $H_{IJK}=3\partial_{[I}B_{JK]}$. 
In the Kaluza-Klein-like manner as 
%
\begin{eqnarray}
G_{MN}dx^Mdx^N=(dy+{\sqrt {2}}A_\mu dx^\mu)^2
+{\hat g}_{\mu\nu}dx^\mu dx^\nu, \label{eq:metric5}
\end{eqnarray}
%
we obtain the four dimensional action
%
\begin{eqnarray}
& & S_4=\int d^4 x{\sqrt {-{\hat g}}} \Bigl[{\hat R}+4{\hat g}^{\mu\nu}
\nabla_\mu \phi \nabla_\nu \phi -\frac{1}{12}{\hat g}^{\mu\nu}
{\hat g}^{\alpha\beta}{\hat g}^{\rho\sigma}
H_{\mu\alpha\rho}H_{\nu\beta\sigma} \nonumber \\
& & ~~~~~~~~~~~~~~~~~~~~~~~~~~~~~~~~~~~
-{\hat g}^{\mu\nu}{\hat g}^{\alpha\beta}
F_{\mu\alpha}F_{\nu\beta}-\Lambda \Bigl]e^{-2\phi},
\end{eqnarray}
%
where $F_{\mu\nu}=\partial_\mu A_\nu -\partial_\nu A_\mu$, 
$H_{\mu\nu\sigma}=3\partial_{[\mu} B_{\nu\sigma]}+24A_{[\mu}
\partial_\nu A_{\sigma]}$ and 
we assumed that all of fields do not depend on the coordinate
$y$. Finally, rewriting the action in the Einstein frame where we 
can move to there by the conformal transformation, 
${\hat g}_{\mu\nu}=e^{2\phi}g_{\mu\nu}$, 
we obtain the action of Eq. (\ref{eq:action1}). 

Inspired by the chiral null mode, we can set  
%
\begin{eqnarray}
{\hat g}_{\mu\nu}dx^\mu dx^\nu=-F^2(d\tau+\omega_i dx^i)^2+\delta_{ij}
dx^i dx^j \label{eq:sol1}
\end{eqnarray}
%
%
\begin{eqnarray}
A=\frac{F}{{\sqrt {2}}}(d\tau +\omega_i dx^i) \label{eq:sol2}
\end{eqnarray}
%
%
\begin{eqnarray}
B_{\mu y}=-{\sqrt {2}}A_\mu \label{eq:sol3}
\end{eqnarray}
%
and
%
\begin{eqnarray}
B_{\mu\nu}=0.\label{eq:sol4}
\end{eqnarray}
%
In fact, in a certain case, the chiral null 
model form of the Lagrangian for the string theory 
becomes\cite{rf:Chiral} 
%
\begin{eqnarray}
L& = & F \partial u [ {\overline \partial}v +F^{-1} {\overline \partial}u
+2\omega_i {\overline \partial}x^i]+\partial x^i {\overline \partial} x_i
\nonumber \\
& = & -F^2(\partial \tau +\omega_i \partial x^i )
({\overline \partial}\tau+\omega_i {\overline \partial}x^i)
+(\partial y+F \partial \tau +F\omega_i \partial x^i)
({\overline \partial}y+F{\overline \partial}\tau+F\omega_i {\overline
\partial}x^i) \nonumber \\
& & +F(\partial y {\overline \partial}\tau - \partial \tau {\overline
\partial}y)+F\omega_i
(\partial y {\overline \partial}x^i - \partial x^i {\overline
\partial}y)+\partial x^i {\overline \partial} x_i,\label{eq:chiral}
\end{eqnarray}
%
where $\partial=\partial_+$ and ${\overline \partial}=\partial_-$. 
From the first to the second and third line, 
we set $u=y$ and $v=2\tau$. Here $\omega_i$ is assumed 
to satisfy $\epsilon_{ijk} \partial_j \omega_k =\partial_i \omega$, 
where $\omega$ a harmonic function.  
Since we are considering the system with a positive cosmological 
constant and this system is not exactly string theory, we remind readers 
that the chiral null model gives just reference to manage the problem. 
In the actual chiral null model $F$ does not depend on the
coordinates $y$ and $\tau$. On the other hand, $F$ does on $\tau$ in
the present study. Comparing the second line of Eq. (\ref{eq:chiral}) with the 
string action, $L=(G_{MN}+B_{MN})\partial X^M {\overline \partial}X^N$, 
we see that Eqs. (\ref{eq:sol1}), (\ref{eq:sol2}), 
(\ref{eq:sol3}) and (\ref{eq:sol4}) together with 
Eq. (\ref{eq:metric5}) are plausible arranging. 

Using Eqs. (\ref{eq:metric5}) and (\ref{eq:sol1}), 
the five dimensional metric of Eq. (\ref{eq:metric5}) becomes 
%
\begin{eqnarray}
G_{MN}dx^Mdx^N=dy^2+2Fdyd\tau+2F\omega_i dydx^i +\delta_{ij}dx^idx^j. 
\end{eqnarray}
%
Equations which we should solve are derived by the variational 
principle of the action (\ref{eq:action2}),
%
\begin{eqnarray}
\partial_I({\sqrt {-G}}e^{-2\phi}H^{IJK} )=0 \label{eq:axion}
\end{eqnarray}
%
%
\begin{eqnarray}
\nabla_M \nabla^M \phi -\nabla_M \phi \nabla^M \phi +\frac{1}{4}
\Bigl(R_G-\frac{1}{12}H^2-\Lambda \Bigr)=0 \label{eq:dilaton}
\end{eqnarray}
%
and
%
\begin{eqnarray}
& & R_{MN} - \frac{1}{2}G_{MN}R_G+2\nabla_M \nabla_N \phi 
-2G_{MN}\nabla_K \nabla^K \phi +2G_{MN}\nabla_K \phi \nabla^K \phi 
\nonumber \\
& &~~~~~~~~~~~~ 
-\frac{1}{4}\Bigl( H_{M}^{~IJ}H_{NIJ}-\frac{1}{6}G_{MN}H^2 \Bigr)
+\frac{1}{2}G_{MN}\Lambda=0. \label{eq:einstein}
\end{eqnarray}
%
Substituting Eq. (\ref{eq:dilaton}) into Eq. (\ref{eq:einstein}) the 
Einstein equation is simplified as  
%
\begin{eqnarray}
R_{MN}+2\nabla_M \nabla_N \phi 
-\frac{1}{4}H_{M}^{~IJ}H_{NIJ} =0.\label{eq:einstein2}
\end{eqnarray}
%
First of all, we assume 
%
\begin{eqnarray}
e^{2\phi}=\frac{F}{a^2},~~a(\tau)=e^{H_0 \tau}
\end{eqnarray}
%
and 
%
\begin{eqnarray}
\partial_\tau ( a^2 F^{-1} ) =2H_0 a^2.
\end{eqnarray}
%
thanks to the dilatonic KT solution\cite{rf:HH,rf:MS}. 
Eq. (\ref{eq:axion}) yields 
%
\begin{eqnarray}
\partial_\tau (a^2 \omega_i)=0 \label{eq:aome}
\end{eqnarray}
%
%
\begin{eqnarray}
\partial_i \partial^i F^{-1}=0\label{eq:harmonic}
\end{eqnarray}
%
and
%
\begin{eqnarray}
\partial_\tau (a^2 \partial_i F^{-1})=0\label{eq:final}
\end{eqnarray}
%
From Eq. (\ref{eq:harmonic}) we can see that 
$F^{-1}$ is a harmonic function. 
Eqs. (\ref{eq:aome})$\sim$ (\ref{eq:final}) 
have the solution with Eqs. (\ref{eq:functionf}) and (\ref{eq:omega}). 
We can check in straightforward way that Eqs. 
(\ref{eq:functionf}) and (\ref{eq:omega})satisfy 
the rest equations (\ref{eq:dilaton}) and (\ref{eq:einstein2}). 
As a result, we obtain the solution of 
Eqs.(\ref{eq:ans1})$\sim$(\ref{eq:ans2}) in the Einstein frame. 

In the same way as the non-rotating 
dilatonic KT solution\cite{rf:HH}, we can see 
that the solution given by us here, the rotating dilatonic KT solution, 
has the timelike singularities which surrounds solitons. 
This is the reason why we say dashed `Black 
Hole' in the title and the abstract of this paper. 
Rigorously speaking, the solution does not 
describe black holes. However, this type might be interpreted as 
a non-singular solution in a higher dimensions as Gibbons, Horowitz
and Townsend have done\cite{rf:GHT}. 

In this Letter, for simplicity, we considered the EMDA-$\Lambda$ theory 
where we can be bearing the chiral null model spirit in mind. 
And now we could find the spinning version of the dilatonic KT solution. 
Thus, we surely expect that the spinning non-dilatonic solution exists 
as well as our present solution. 
We may think that the spinning version of the KT solution in 
the EM-$\Lambda$ theory 
is free from naked singularities because the KT solution in the 
pure Einstein-Maxwell theory is like that. Aided the study of the 
force balance up to the gravitational spin-spin 
interaction\cite{rf:Gen}, we think that 
the present work can encourage us to find a spinning version of the 
KT solution in the EM-$\Lambda$ theory. Hopefully, we would like to 
discover the exact solution in near future. At the same time, 
the comprehensive features such as the global structure and the
symmetry like supersymmetry should be reported together with the
features of the present rotating dilatonic KT solution if there is.  

\section*{Acknowledgements}
The author would like to thank Gary W. Gibbons and DAMTP 
relativity group for their hospitality at Cambridge. 
This work is supported by JSPS Postdoctal Fellowship for 
Research Abroad. 

%


\begin{thebibliography}{99}
\bibitem{rf:KT}
D.~Kastor and J.~Traschen, \PR{D47,1993,5370}
\bibitem{rf:Brill}
D. R. Brill, G. T. Horowitz, D. Kastor and J. Traschen, \PR{D49,1994,840}
\bibitem{rf:Nakao}
K. Nakao, T. Shiromizu and S. A. Hayward, \PR{D52,1995,796};\\
D. Ida, K. Nakao, M. Siino and S. A. Hayward, \PR{D58,1998,121501} 
\bibitem{rf:nohair}
G. W. Gibbons and S. W. Hawking, \PR {D15,1977,2738};\\
S. W. Hawking and I. G. Moss, \JL{Phys.~Lett.,110B,1982,35};\\
T. Shiromizu, K. Nakao, K. Maeda and H. Kodama, \PR{D47,1993,R3099};\\
S. A. Hayward, T. Shiromizu and K. Nakao, \PR{D49,1994,5080};\\
K. Maeda, T. Koike, M. Narita and A. Ishibashi, \PR{D57,1998,3503} 
\bibitem{rf:inf}
A.~H.~Guth, \PR{D23,1981,347};\\
K.~Sato, \JL{Mon.~Not.~R.~Astron.,195,1981,467};\\
A.~D.~Linde, {\it Particle Physics and Inflationary
Cosmology}(Harwood,~Switzerland, 1990)
\bibitem{rf:IW}
W.~Israel and G.~A.~Wilson, \JMP{13,1972,865}
\bibitem{rf:Perjes}
Z.~Perjes, \PRL{27,1971,1668}
\bibitem{rf:Force}
J.~Tiomno, \PR{D7,1973,356};\\
W.~Israel and J.~T.~J.~Spanos, Lett. \NC{7,1973,245};\\
D.~Kastor and J.~Traschen, Class. Quantum Grav. {\bf 16} (1999),1265
\bibitem{rf:Gen}
T.~Shiromizu and U.~Gen, Report No. DAMTP-1999-100, submitted 
\bibitem{rf:Chiral}
G.~T.~Horowitz and A.~A.~Tseytlin, \PR{D51,1995,2896}
\bibitem{rf:SUSY}
R.~Kallosh, D.~Kastor, T.~Ortin and T.~Torma, \PR{D50,1994,6374}
\bibitem{rf:Tess1}
T.~Shiromizu, hep-th/9907017, to be published in Phys.~Rev. {\bf D}
\bibitem{rf:HH}
J.~H.~Horne and G.~T.~Horowitz, \PR{D48,1993,5457}
\bibitem{rf:MS}
T.~Maki and K.~Shiraishi, Class. Quantum Grav. {\bf 10} (1993), 2171
\bibitem{rf:GHT}
G.~W.~Gibbons,~G.~T.~Horowitz and P.~K.~Townsend,
\JL{Class.~Quantum~Grav.,12,1995,297}
\end{thebibliography}
\end{document}